\documentclass[aps,pre,floatfix,twocolumn,showpacs,showkeys,amsmath,amssymb,superscriptaddress,10pt]{revtex4-1}

\usepackage{amsthm}
\usepackage{bm}
\usepackage{epsfig}
\usepackage{verbatim}
\usepackage{graphicx}

\newcommand{\beq}{\begin{equation}}
\newcommand{\eeq}{\end{equation}}
\newcommand{\beqa}{\begin{eqnarray}}
\newcommand{\eeqa}{\end{eqnarray}}

\usepackage{xcolor}

\begin{document}

\title{Modulated phases in external fields: when is reentrant behavior  to be expected?}

\author{Alejandro Mendoza-Coto}
\affiliation{Departamento de F\'\i sica,
Universidade Federal do Rio Grande do Sul,
CP 15051, 91501-970, Porto Alegre, Brazil}
\author{Orlando V. Billoni}
\email{billoni@famaf.unc.edu.ar}
\affiliation{Facultad de
Matem\'atica, Astronom\'{\i}a, F\'{\i}sica y Computaci\'on, Universidad Nacional
de C\'ordoba, Instituto de F\'{\i}sica Enrique Gaviola (IFEG-CONICET)\\ Ciudad Universitaria, 5000 C\'ordoba, Argentina}
\author{Sergio A. Cannas}
\email{cannas@famaf.unc.edu.ar}
\affiliation{Facultad de
Matem\'atica, Astronom\'{\i}a, F\'{\i}sica y Computaci\'on, Universidad Nacional
de C\'ordoba, Instituto de F\'{\i}sica Enrique Gaviola (IFEG-CONICET)\\ Ciudad Universitaria, 5000 C\'ordoba, Argentina}
\author{Daniel A. Stariolo}
\email{stariolo@if.uff.br}
\affiliation{Departamento de F\'{\i}sica,
Universidade Federal Fluminense and
National Institute of Science and Technology for Complex Systems\\
Av. Gal. Milton Tavares de Souza s/n, Campus Praia Vermela,
24210-346 Niter\'oi, RJ, Brazil}

\date{\today}

\begin{abstract}
We introduce a new coarse grain model capable of describing the phase behavior of two dimensional ferromagnetic systems
with competing exchange and dipolar interactions, as well as an external magnetic field. An improved expression for the mean field
entropic contribution allows to compute the phase diagram in the whole temperature versus external field plane. We find
that the topology of the phase diagram may be qualitatively different depending on the ratio between the strength of the
competing interactions. In the regime relevant for ultrathin ferromagnetic films with perpendicular anisotropy we confirm
the presence of inverse symmetry breaking from a modulated phase to a homogenous one as the temperature is lowered at
constant magnetic field, as reported in experiments. For other values of the competing interactions we show that reentrance
may be absent. Comparing thermodynamic quantities in both cases, as well as the evolution of magnetization profiles in the
modulated phases, we conclude that the reentrant behavior is a consequence of the suppression of domain wall degrees of
freedom at low temperatures at constant fields.
\end{abstract}


\maketitle

\section{Introduction}
\label{Intro}
The field versus temperature phase diagram of ultra-thin ferromagnetic films displaying stripe, bubbles
and homogeneous phases has attracted attention in recent years, mainly due to the existence of
new experimental results from which the phase diagram and other interesting characteristics of the
phase transitions have been reported~\cite{SaLiPo2010,SaRaViPe2010}. Early theoretical results on the
phase diagram from an effective model with dipolar interactions were challenged~\cite{GaDo1982} by
experiments.
The main qualitative difference between early phase diagrams and recent experimental results was the
observation of an inverse symmetry breaking (ISB) transition, with a sequence of homogeneous-modulated-homogeneous phases, as
the temperature is lowered at fixed external field~\cite{SaLiPo2010,SaRaViPe2010}.  The existence of an ISB transition in such systems had been predicted in the pioneer work of Abanov et al\cite{AbKaPoSa1995}, using a phenomenological approach.  Subsequent theoretical
work analyzed the existence of ISB from a scaling hypothesis~\cite{PoGoSaBiPeVi2010,MeSt2012}.
Reentrant behavior was shown on a coarse-grained model of the Landau-Ginzburg type~\cite{CaCaBiSt2011},
although no attempt was made to explain the nature of the
reentrance, mainly due to limitations in the very definition of the model, which was not able to capture
the low temperature sector of the phase diagram. Recently, Velasque {\it et. al.} studied a mean field
version of the dipolar frustrated Ising ferromagnet (DFIF), and showed that the stripe phase in a field presents
reentrant behavior~\cite{VeStBi2014}. Furthermore, by comparing the DFIF with two simpler models, the
authors concluded that the reentrant behavior in this kind of systems has its origin on the entropy gain
from domain walls degrees of freedom of modulated structures.

Inverse freezing in magnetic models has been observed mainly in spin glasses and disordered
systems, in which frustration leads to complex entropic contributions~\cite{Se2006,PaLeCr2010,SiZiMaLa2012,ErThMa2013}.
Nevertheless, the question of the physical origin of reentrant behavior remains obscured by the inherent
complexity of the thermodynamic behavior of disordered systems. Magnetically frustrated systems without
quenched disorder, where low temperature phases and ground states display known symmetries,
 seem to be better candidates for getting a better understanding of ISB~\cite{PoGoSaBiPeVi2010,CaCaBiSt2011,VeStBi2014}.
Besides ultrathin ferromagnetic systems with dipolar frustration, other frustrated systems without quenched disorder
showing inverse transitions are, e.g. the $J_1$-$J_2$ model in
the square lattice~\cite{Yin09,Queiroz13} and the axial next-nearest-neighbor Ising (ANNNI) model~\cite{YoCoSa1981}.

The aim of the present work is twofold: first, we introduce a new coarse-grain model for ultrathin
ferromagnetic films with perpendicular anisotropy which, at variance with previous ones,
is valid at any temperature, allowing the computation of the complete phase diagram.
By minimizing the corresponding free energy in a mean field approximation, we obtain the magnetic field
versus temperature phase
diagram showing homogeneous paramagnetic, stripes and bubbles phases. This spans the complete phenomenology
observed in experiments~\cite{SaLiPo2010,SaRaViPe2010}. Furthermore, we show that in the experimentally relevant sector of coupling constants
the system shows inverse symmetry breaking, but for general values of the ratio between the competing interactions,
this is not always the case. Thus, we concentrate our discussions on two relevant cases, one showing ISB and another
without reentrance, and analyze the origin of the different behaviors between them.
Second, we discuss the physical origins of inverse symmetry breaking in this kind of
systems. We present compelling evidence that the inverse transition is driven by the excess of
degrees of freedom present in the domain walls, namely, a basically entropy-driven mechanism.
We show that the domain wall structure and evolution with
temperature and magnetic field is very different in the two prototypical cases studied, which reinforces the
argument on the relevance of domain wall structure for ISB, complementing and expanding the analysis of
reference \onlinecite{VeStBi2014}.

The organization of the paper is as follows: in Section \ref{model} we introduce the model and the mean field
approximation.
In \ref{phases} we present the results for the magnetic field versus temperature phase diagrams and analyze the nature of
the reentrant behavior and the nature of the phase transitions observed.
In Section \ref{conc} we conclude with a summary of the results.

\section{Model}
\label{model}
Modulated phases in ultrathin ferromagnetic films occur at mesoscopic scales, i.e. the typical length scale of
the modulations in the magnetization density are much larger than the lattice spacing, thus justifying a coarse grained
description (see, e.g. reference \cite{PoGoSaBiPeVi2010} for a detailed justification of the coarse grained description
in this kind of systems). Then, our starting point is the effective hamiltonian:
\begin{widetext}
\beq
H[\phi]=\frac{1}{2}\int d^2x\; \left(\vec \nabla \phi(\vec x) \right)^2 +\frac{1}{2} \iint d^2x d^2x' \,J(|\vec x-\vec x'|)
\phi({\vec x})\phi({\vec x'}) - B\,\int d^2x\; \phi({\vec x}),
\label{effH}
\eeq
\end{widetext}
\noindent where $\phi({\vec x})$ is the out of plane magnetization density, the first term represents the effective short range exchange interaction,
the second term is a competing long range
dipolar interacion, which in the limit of strong perpendicular anisotropy reduces to the form $J(x)= J/x^3$ and the last one is a coupling to an external homogeneous magnetic field $B$
perpendicular to the plane of the film. Within a mean field approximation, we can then construct an effective free energy functional $F[\phi]= H[\phi]-T\, S[\phi]$, which after a minimization respect to the field $\phi$ gives us the equilibrium state ($S[\phi]$ being some properly defined entropy functional). Then, the effective free energy reads
\begin{widetext}
\begin{equation}
F[\phi] = \frac{1}{2} \iint d^2x d^2x' \,A(\left|{\vec x}-{\vec x}'\right|)\phi({\vec x})\phi({\vec x'})
- \frac{1}{\beta} \int d^2x\; {\cal S}(\phi({\vec x})/\phi_0) -
B\,\int d^2x\; \phi({\vec x}) \label{Hreal}
\end{equation}
\end{widetext}
\noindent where ${\cal S}(x)$ is an entropy density, $\phi_0$ corresponds to the saturation value of the magnetization, $\beta=(k_BT)^{-1}$
and the quadratic kernel $A(\left|{\vec x}-{\vec x}'\right|)$ encodes all the information about the physical interactions in the system.
Note that, up to this point, the model defined is quite general.
Previously considered coarse grained models have been mainly of the Ginzburg-Landau type, defined by an expansion in powers of the order parameter,
typically up to $\phi^4$, which limits the validity of results to temperatures near the critical point~\cite{PoGoSaBiPeVi2010,CaCaBiSt2011}.
Instead, in line with the mean field approximation, we consider
an entropy density function to be of the form:
\begin{equation}\label{entropy1}
{\cal S}(x) = \frac{1+x}{2}\log{\frac{1+x}{2}}+\frac{1-x}{2}\log{\frac{1-x}{2}}.
\end{equation}
This form imposes saturation values to the order parameter $|\phi(\vec x)| \leq \phi_0$ and allows a computation of the thermodynamic
properties for any temperature.

It is well known that the solutions which minimize the effective free energy (\ref{Hreal}) (at low enough temperatures) correspond to periodic patterns in space
in the form of stripes or bubbles~\cite{GaDo1982,SeAn1995}.
The general solution for the order parameter can be written as a Fourier series expansion
of the form $\phi(\vec{x})=\sum_{i=0}^{\infty}c_i\cos\left(\vec{k}_i\cdot\vec{x}\right)$.  Different sets of
wave vectors will define different patterns, so part of the problem is to choose the appropriate set of wave vectors for constructing
the particular solutions expected. Replacing the general solution into Eq.(\ref{Hreal}),  and
after a Fourier transformation, the free energy density reads:
\begin{widetext}
\beq
f[\phi]=\frac{F[\phi]}{V} =\frac{1}{2}\hat{A}(0)\,c_0^2+\frac{1}{4}\sum_i \hat{A}(k_i)\,c_i^2
-\frac{1}{\beta V}\int d^2x\; {\cal S}\left(\sum_ic_i\cos\left(\vec{k}_i\cdot\vec{x}\right)\right)-B\, c_0 ,
\label{Ham1}
\eeq
\end{widetext}
where $V$ is the total volume (area) of the system. The function $\hat{A}(\vec{k})$  stands for the Fourier
transform of $A(x)$ (fluctuation spectrum) and $c_0$ represents the amplitude of the zero wave vector mode. Since the function $A(x)$ is the
sum of two competing interactions, it turns out that the Fourier transform of the model defined in (\ref{effH}) has a minimum at a nonzero
wave vector $k_0$. This signals the fact that the competition between exchange and dipolar interactions favors the formation
of periodic patterns in the order parameter. The value of $k_0$  corresponds to the optimum wave vector for the formation of single mode modulated
structures and sets a natural characteristic length scale for the system. Hence, from now on, all wave vectors will be expressed in units of $k_0$ ($k_0=1$)
and all lengths  in units of $2\pi/k_0$.

It is also useful to express the energy in units of  $\vert\hat{A}(k_0)\vert$ and the temperature in units of $\vert\hat{A}(k_0)\vert/k_B$, so
that
\beq
f[\phi] =\frac{1}{2}{\cal A}(0)\,c_0^2+\frac{1}{4}\sum_i {\cal A}(k_i)\,c_i^2
-Ts-h\,c_0 ,
\label{Ham2}
\eeq

\noindent where ${\cal A}(k)=\hat{A}(k)/\vert\hat{A}(k_0)\vert$,
$h=B/\vert\hat{A}(k_0)\vert$ and $s$ is the entropy per volume unit. The expression above
is written in terms of dimensionless variables only, which makes it suitable for numerical work.

In order to develop modulated patterns of typical scale $k_0$, $A(k)$ should have a
negative minimum. Such condition ensures the necessary stability of those modes near the circumference
of radius $k_0$, and consequently the formation of modulations in the order parameter.
Consequently  ${\cal A}(k_0)=-1$.
Considering that the short range part of $A(k)$ is proportional to $k^2$ in the long wavelength limit, and that in the same limit the
dipolar interaction gives a contribution proportional to $-k$~\cite{CzVi1989,PiCa2007}, the appropriate form for ${\cal A}$
for the systems considered here has the general form:
\beq
{\cal A}(k)=-1+a(k-1)^2.
\label{spectrum}
\eeq

\begin{figure}[ht!]
\includegraphics[scale=0.35]{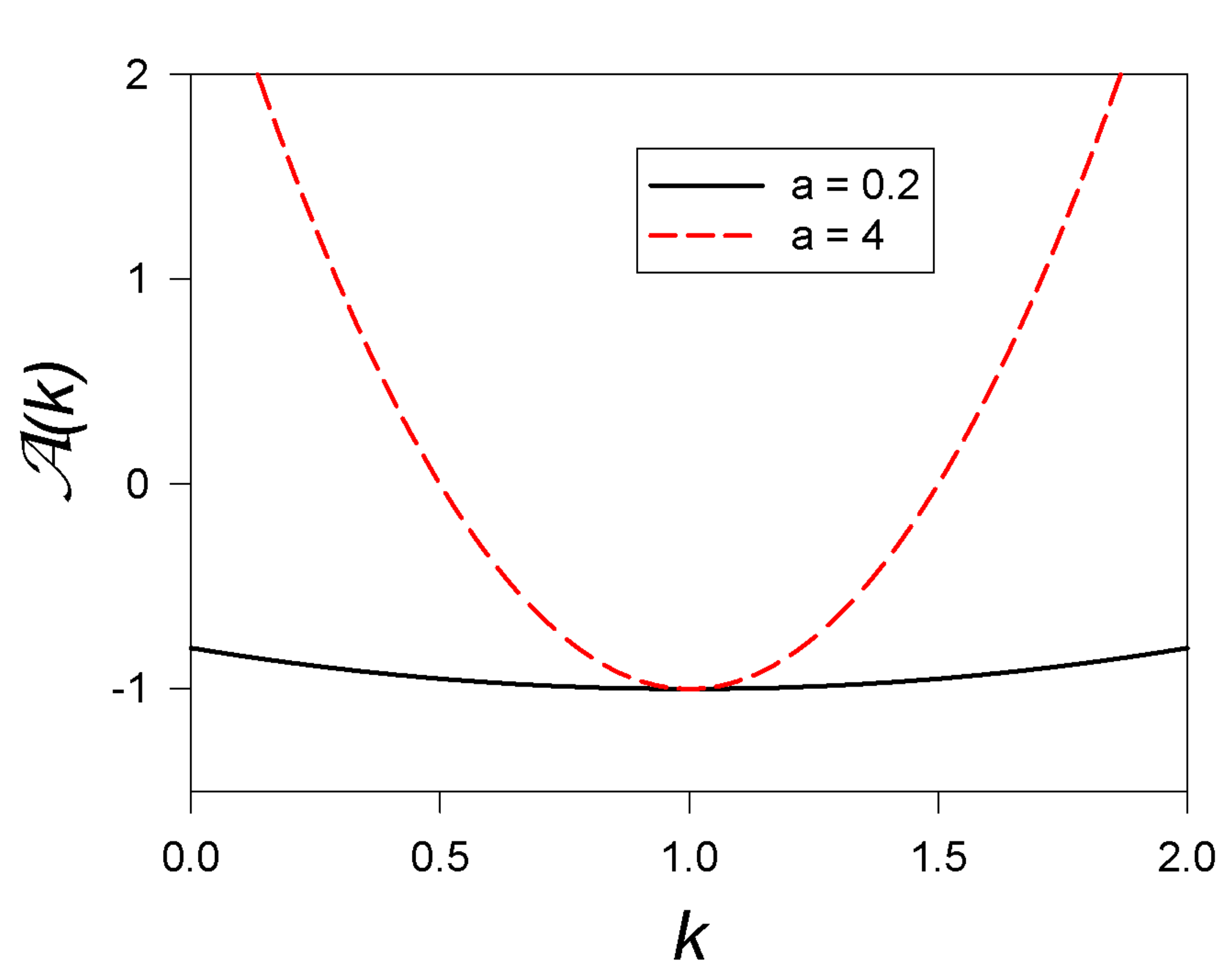}
\caption{(Color online) The spectrum of fluctuations $A(k)$ of equation (\ref{spectrum}) for two values of the curvature parameter $a=0.2$ and $a=4$. }
\label{fig:spectra}
\end{figure}

The only free parameter in the fluctuation spectrum (\ref{spectrum}) is the curvature $a$. In figure \ref{fig:spectra}
the fluctuation spectrum is shown for two representative values of the parameter $a$. In the following it will be shown that
these two cases have very different phase diagrams.

Regarding the ground states or low energy configurations of these kind of systems, to our knowledge, the only exact results available
correspond to the ground states of the square lattice Ising model with ferromagnetic nearest neighbor interactions plus antiferromagnetic dipolar
interactions proportional to $1/r^3$~\cite{GiLeLi2006,GiLeLi2011}, where it has been shown that the ground states at zero external field
are striped patterns for large enough ferromagnetic interactions, limit relevant to experimental ultrathin films with perpendicular anisotropy.
For finite external fields there are no exact results but, based on experimental evidence of low temperature patterns, a series of interesting works have
compared the energetics of striped, bubbles, checkerboard and homogeneous configurations~\cite{CaLe1971,NgVa1995,SaRaViPe2010}. All the theoretical
evidence indicates that, at zero temperature and low enough fields, the striped configurations have the lower energy, until a critical field value
from where an hexagonal array of bubbles becomes the ground state. At still a higher critical field, the homogeneously magnetized state turns to be
the lowest energy state.
Then, the relevant equilibrium configurations of the density field $\phi(\vec x)$  may be of two different types\footnote{Of course, more realistic patterns are observed in
experiments or obtained in computational simulations. The perfect
stripes or bubbles patterns considered here are mean field approximations to the real solutions, in the sense that fluctuations and defects are
not taken into account. Nevertheless, they represent a good qualitative approximation to the real situation.}:
striped configurations which can be written in the form:

\begin{equation}
    \phi_s({\vec x}) =  \sum_{i=0}^\infty c_i \cos (k_{eq}\, \vec{s}_i\cdot{\vec x}),
\label{stripes}
\end{equation}

\noindent with the vectors ${\vec{s}_i}=i(1,0)$, and bubble configurations:

 \begin{equation}
    \phi_b({\vec x}) =  \sum_{i=0}^\infty c_{i} \cos (k_{eq}\, {\vec{b}}_i\cdot{\vec x}),
\label{bubbles}
\end{equation}

\noindent where the set of vectors ${\vec{b}}_i$ are defined on a triangular lattice with lattice spacing equal to one.
The details of the definitions of the wave vectors forming the bubbles solutions are shown in the Appendix.
In both cases
$k_{eq}$ represents the equilibrium wave vector that defines the modulation length for each structure.
Substituting Eqs.(\ref{stripes}) and (\ref{bubbles}) into Eq.(\ref{Ham2}) leads to a mean field variational free energy in terms of the infinite
set of amplitudes $\{ c_n\}$ and $k_{eq}$. After truncating Eqs.(\ref{stripes}) and (\ref{bubbles}) to some maximum number of modes $n_{max}$,
 variational expressions at different levels of approximation for the stripes and bubbles free energies are obtained.
Assuming that the only equilibrium states are stripes, bubbles or homogeneous ones, we determined the equilibrium phase diagram
by minimizing and comparing the free energies for each type of solution to the same fixed level of approximation $n_{max}$. The functional
minimization was performed by the method of Gaussian quadratures.

\section{Results}
\label{phases}
\begin{figure}[ht!]
\includegraphics[scale=0.3,angle=-90]{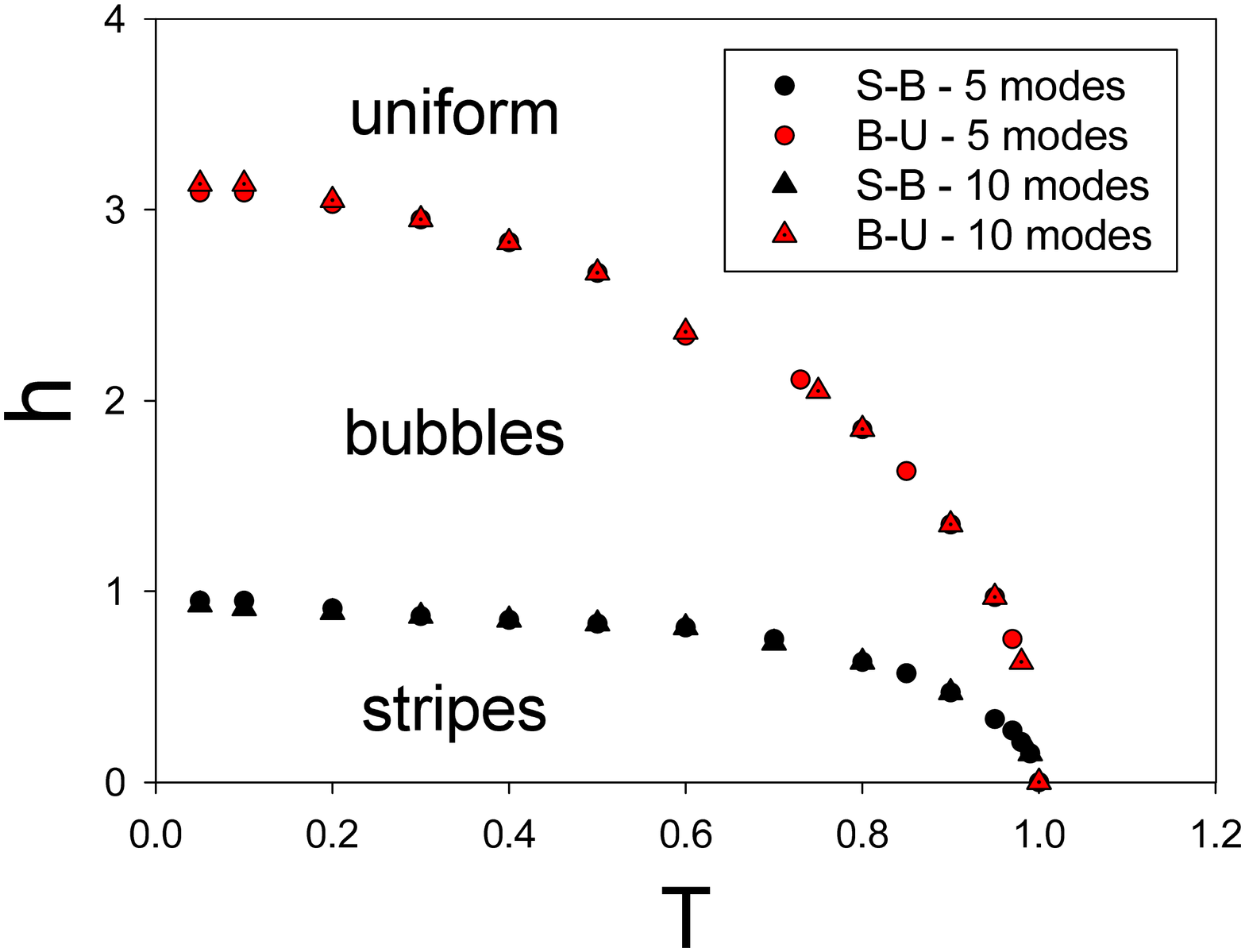}
\caption{(Color online) External field versus temperature phase diagram for $a=4$ and different degree of approximations $n_{max}$. }
\label{fig:pda4}
\end{figure}

\begin{figure}[ht!]
\includegraphics[scale=0.5]{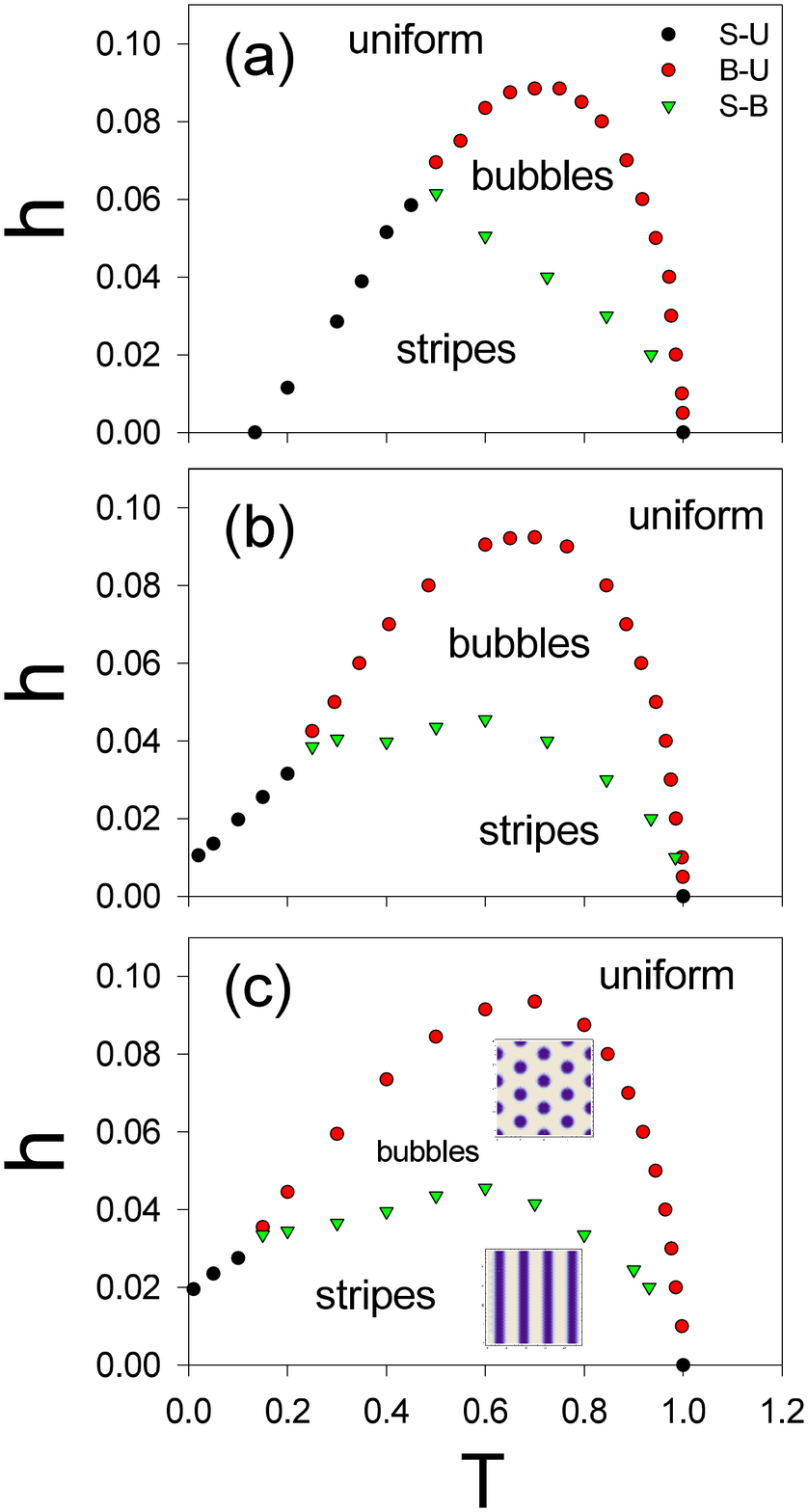}
\caption{(Color online) External field versus temperature phase diagram for $a=0.2$. Different panels
correspond to $n_{max}$ equal to (a) 5, (b) 10, (c) 15.}
\label{fig:pda02}
\end{figure}

In figures \ref{fig:pda4} and \ref{fig:pda02} we show the  magnetic field ($h$) versus temperature ($T$) phase
diagrams for two representative cases: $a=4$ and $a=0.2$. As can be seen, the topology is very different in each case.
In both cases, three thermodynamic phases can be possible: stripes, bubbles and uniform, qualitatively similar to observations in experiments.
The value of $a$ determines whether the phase diagram will show
reentrant behavior or not, both cases being possible. Comparing expression (\ref{spectrum}) with a more microscopic one, e.g.
the spectrum of the dipolar frustrated Ising ferromagnet considered in \cite{PiCa2007,VeStBi2014}, it can be shown that  $a \propto 1/\delta^2$, where $\delta=J/g$,
 $J$ being  the strength of the short range exchange interaction and $g$  the intensity of the competing dipolar interaction. In the modulated sector of the dipolar frustrated Ising model, $\delta \gg 1$ and therefore $a$ takes typically a small value. We will see in the
following that this leads to reentrant behavior, consistent with what is observed in experiments on ultrathin ferromagnetic films.
Nevertheless, for small $\delta$, reentrant behavior is absent, as can be seen in the phase diagram of figure \ref{fig:pda4}.

As shown in the figure, in this case at low fixed temperature the model goes through two successive transitions as the external field is raised. At zero and
low fields the stripe configurations are the equilibrium phase of the model~\cite{PiCa2007,CaCaBiSt2011,VeStBi2014}, but at a critical field the magnetized background
triggers an instability towards bubble solutions, which are the equilibrium ones at intermediate fields until a transition to a
uniformly magnetized paramagnetic state takes place at a second critical field. As the field is raised the stripes develop a
finite magnetization in the form of an asymmetry favoring the direction parallel to the field. This asymmetry grows and eventually leads
to the bubble equilibrium phase. Both the asymmetry and modulation length grow with the field in a way similar to what was observed in the dipolar
frustrated Ising model in Ref.[\onlinecite{VeStBi2014}], and  seems to diverge at the critical field where the homogenous phase sets in, as it will be discussed later. Remarkably, both transition lines become almost independent of $n_{max}$ for relatively small values of it ($n_{max}=5$) even at very small temperatures. This means that both modulated solutions present basically the same wall structure at all temperatures.

For small values of $a$, the $h-T$ pase diagram shows reentrant behavior. In figure \ref{fig:pda02} we show three instances of the phase
diagram, depending on the maximum number of modes considered in the stripes and bubbles solutions given by equations (\ref{stripes}) and (\ref{bubbles}).
In the bottom panel we show two pictures illustrating the real space stripes and bubbles solutions. For the three
cases considered, $n_{max}=5,\,10$ and $15$, the qualitative picture is the same. Nevertheless, it can be seen that the triple point, at which the three different phases
meet, drifts to the left as $n_{max}$ grows. Then, it can be expected that, in the limit $n_{max}\to \infty$,  this point will either be at $T=0$ or it will simply disappear. Computational limitations
prevented us of reaching larger values of $n_{max}$, and then the solutions began to be unreliable at very low temperatures, where more and more modes acquire a finite
weight in the variational profiles. This trend is in line with experimental results on the phase diagram for Fe/Cu(001) ultrathin films~\cite{SaLiPo2010,SaRaViPe2010}.

\begin{figure}[ht!]
\includegraphics[scale=0.45]{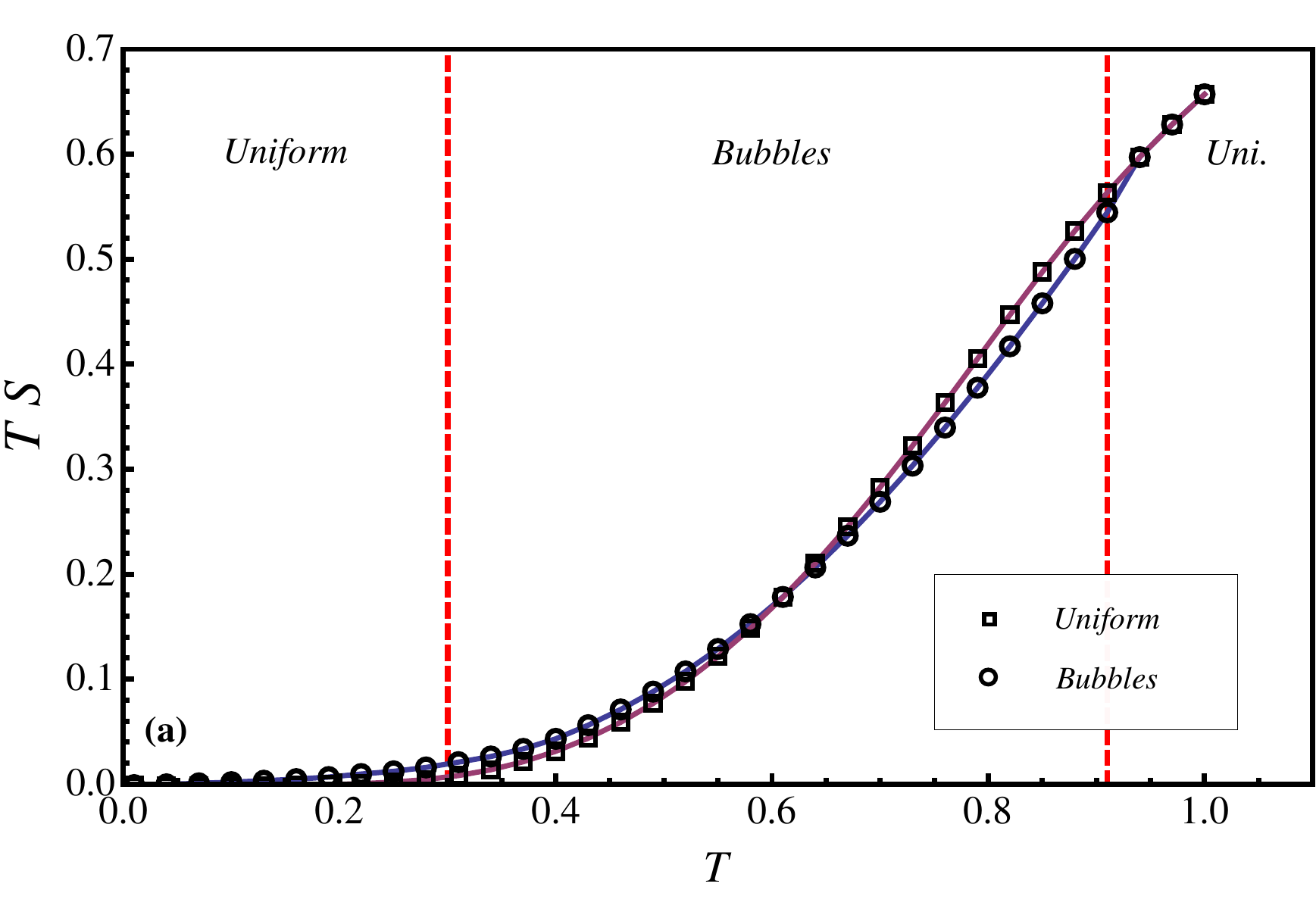}
\includegraphics[scale=0.45]{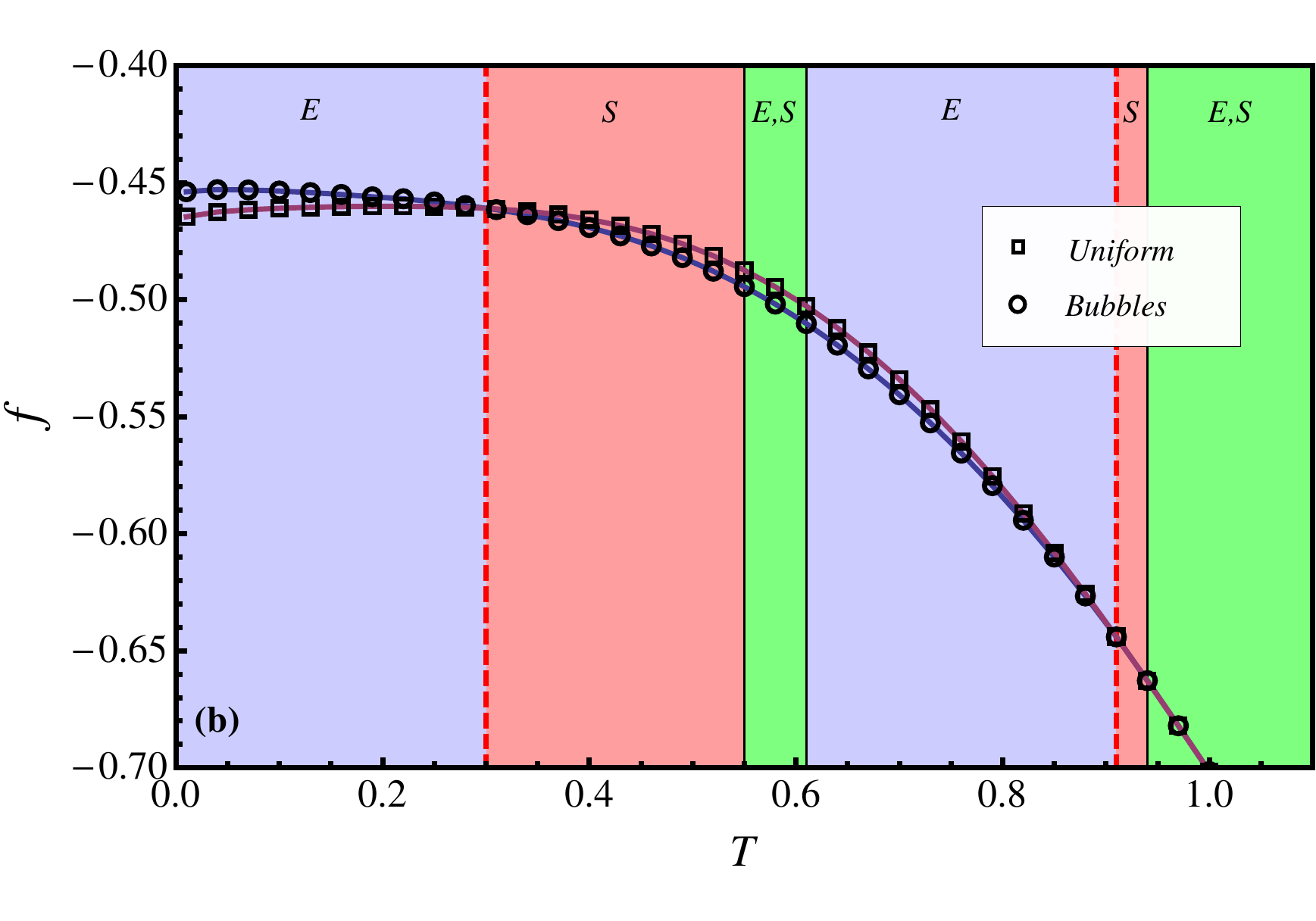}
\caption{(Color online) Thermodynamic functions of the bubbles and uniform phases for $a=0.2$, $h=0.06$ and $n_{max}=15$.  The dashed red lines mark the different transition temperatures for the present field. (a) $T\,S$ as a function of $T$.  (b) Free energy  as a function of $T$. The background colors indicate whether the free energy balance favors the stable phase as a result of having both smaller internal energy  and larger entropy than the other (E,S), larger entropy only (S) or  smaller energy only (E).}
\label{fig:termoa02}
\end{figure}

\begin{figure}[ht!]
\includegraphics[scale=0.45]{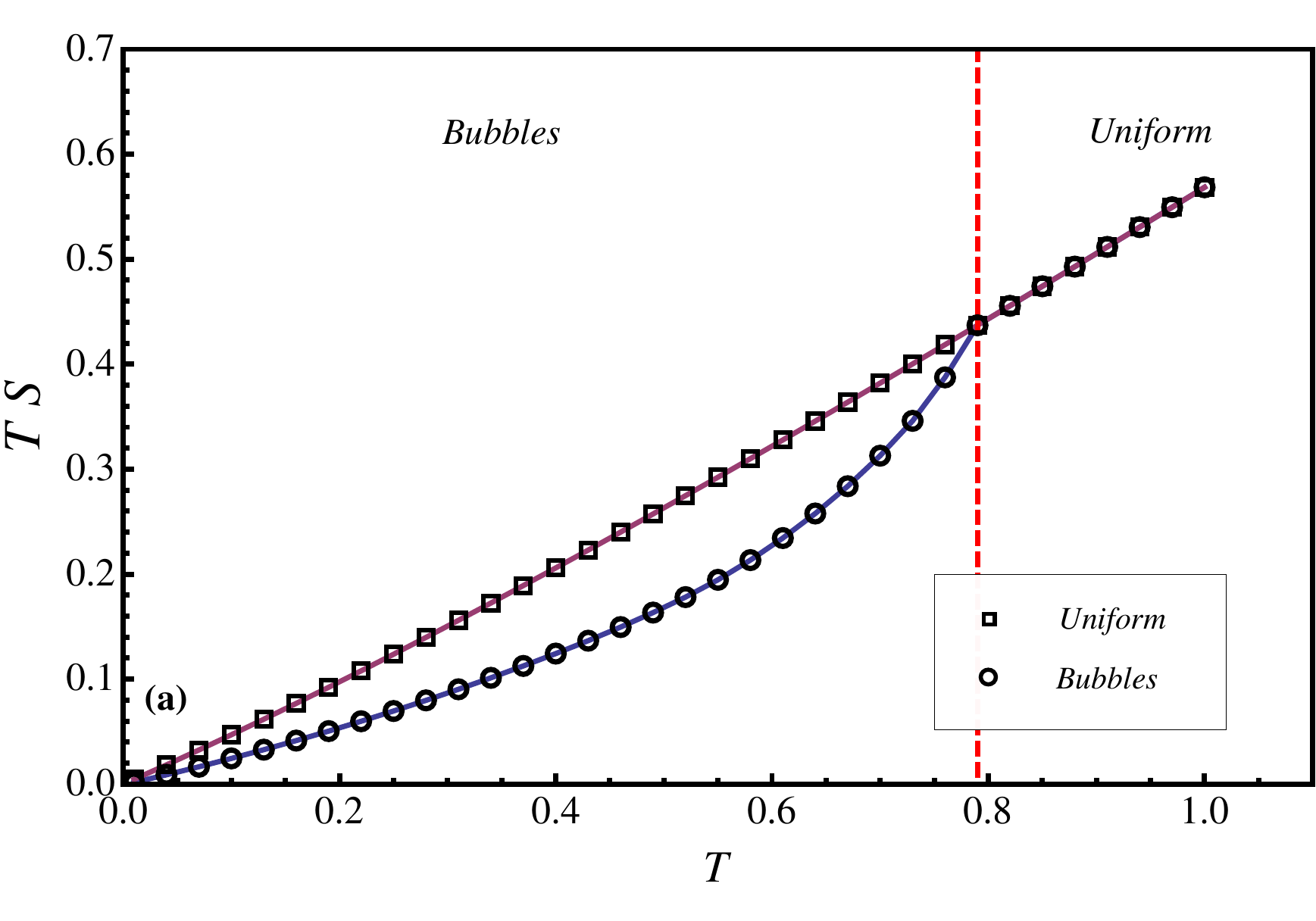}
\includegraphics[scale=0.45]{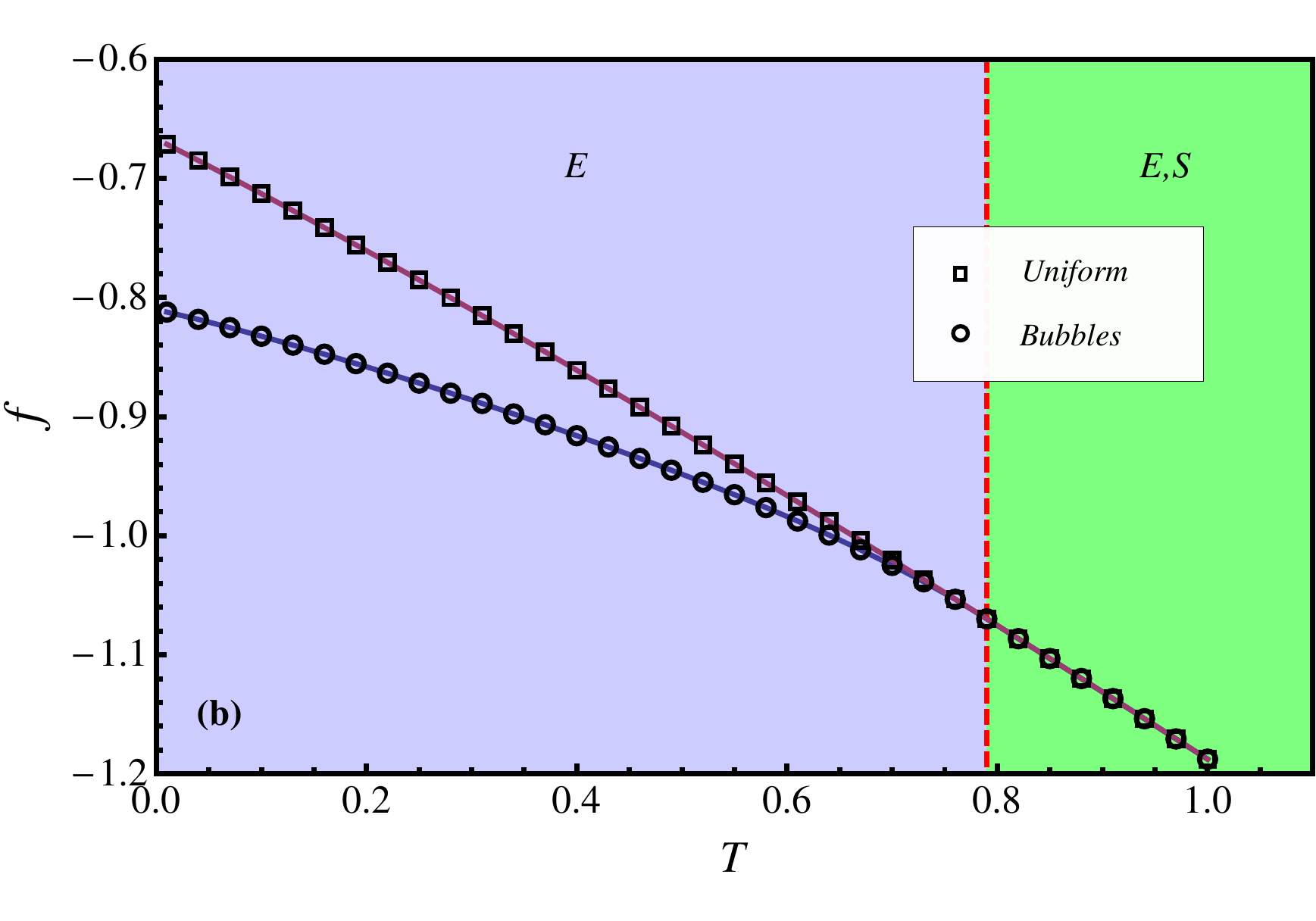}
\caption{(Color online) Thermodynamic functions of the bubbles and uniform phases for $a=4$, $h=2$ and $n_{max}=15$.   The dashed red lines mark the transition temperature for the present field(a) $T\,S$ as a function of $T$.  (b) Free energy  as a function of $T$. The background colors indicate whether the free energy balance favors the stable phase as a result of having both smaller internal energy  and larger entropy than the other (E,S) or just smaller energy (E).}
\label{fig:termoa4}
\end{figure}

To get some insight on the nature of the reentrant behavior we first look at the thermodynamic functions. Figure \ref{fig:termoa02} shows
the entropy and free energy
versus temperature, for $a=0.2$ and fixed external field $h=0.06$, which goes through the bubbles phase for $n_{max}=15$ (see figure \ref{fig:pda02}). The free energy crossing associated with both phase transitions (uniform to bubbles and bubbles to uniform as $T$ increases) appear clearly in Fig.\ref{fig:termoa02}b. We see that both phase transitions result from a subtle balance of both the energy $E$ and the entropy $S$ ($f=E-TS$) as indicated by color codes in the figure. At temperatures $T < 0.55$ both the energy (not shown) and the entropy of the bubbles phase are larger than the corresponding quantities in the uniform phase. At very low temperatures the influence of the energy is stronger than the influence of the entropy and the uniform phase is the stable one. At the ISB transition point ($T\approx 0.3$) such balance inverts and the bubbles phase becomes the stable one. Hence, the behavior of the entropy becomes crucial to the appearance of the ISB transition. On the other hand, at the direct transition from bubbles to uniform ($T\approx 0.9$) the roles played by energy and entropy are interchanged: the uniform phase has both larger entropy and energy than the bubbles. At the transition point the decrease in the internal energy of the uniform phase counterbalances the larger entropy of the bubbles. Consistently, the same behavior is observed in Fig.\ref{fig:termoa4} when $a=4$, in the non reentrant regime. In this case, the entropy of the uniform phase is larger than in the bubbles one at {\it all} temperatures, so the only relevant thermodynamical quantity is the energy, at least as far as the phase transition is concerned.

\begin{figure}[ht!]
\includegraphics[scale=0.47]{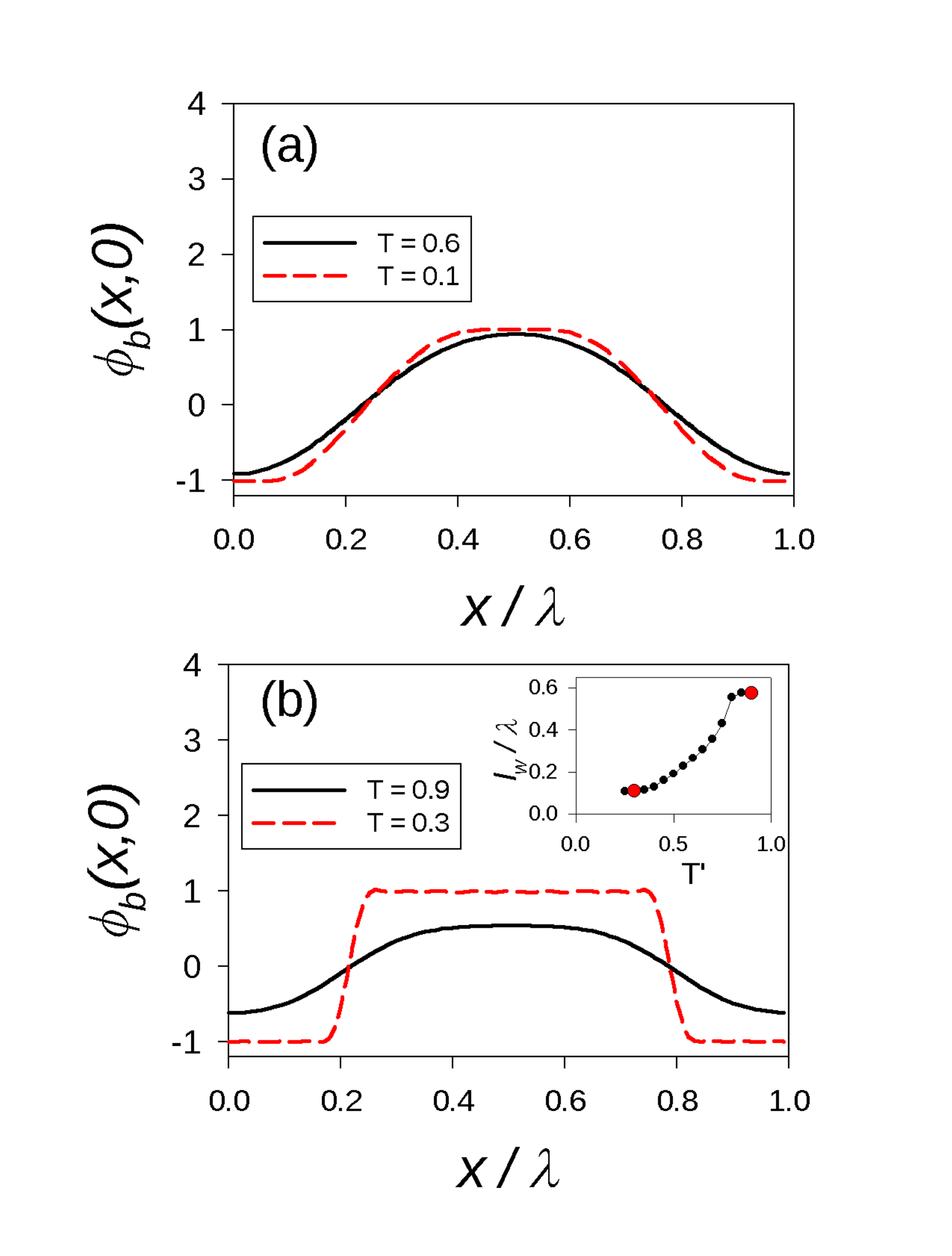}
\caption{(Color online) Domain wall width behavior in the  bubbles magnetization profiles for $n_{max}=15$.  (a) Non reentrant region $a=4$ and $h=2$.
(b)  Reentrant region $a=0.2$ and $h=0.06$. The inset shows the domain wall width $l_w$ to wave length $\lambda$ ratio ($\lambda=2\pi/k_{eq}$) as a
function of the temperature. Red circles mark the boundaries of the bubbles phase. }
\label{fig:profiles}
\end{figure}

In reference [\onlinecite{VeStBi2014}] it was suggested that the excess entropy of domain walls degrees of freedom was responsible for the
reentrant behavior seen in the dipolar frustrated Ising model in the stripes phase. Here, we confirm that expectation. It is
again instructive to compare the
behavior of characteristic quantities for the cases with and without reentrance in our model.
In figure \ref{fig:profiles} we show two dimensional cuts of the bubbles magnetization profiles in the two cases $a=4$ (top panel)
and $a=0.2$ (bottom panel) at a fixed
value of the magnetic field, characteristic in each phase diagram, and $n_{max}=15$. In each panel two profiles, corresponding to two
characteristic temperatures, are
shown. The temperatures were chosen to be near the high temperature transition and a sufficient low temperature in each case. It is
possible to see that the profiles
change little between both temperatures in the model without reentrance. The domain wall width is large, of the order of the modulation
length $\lambda=2\pi/k_{eq}$, both at high and low
$T$. On the contrary, in the model with ISB the profile changes qualitatively from high to low $T$. At high $T$ the profile is more
sinusoidal; few modes have finite
weight in the Fourier expansion. As the temperature is lowered more and more modes acquire a finite weight, and the profile evolves to a
square wave-like one. At exactly
zero temperature, the profile will be exactly a square wave, which has zero entropy. From the entropic contribution perspective, this
means that while for the $a=4$ model the entropy contribution
of domain walls is nearly the same in the whole temperature range, for the $a=0.2$ model the walls rapidly loose entropy as the
temperature approaches low values. In the
inset of the bottom panel in figure \ref{fig:profiles} we show the change in domain wall width relative to the modulation length. Our
conclusion is that this important
loss of entropy of domain walls is responsible for the inverse symmetry breaking phenomenon observed in ultrathin ferromagnetic films.
Also note that, according to figure \ref{fig:spectra}, for large values of
the inverse curvature $a$, the low energy physics is dominated by a few modes around the minimum of Eq.(\ref{spectrum}). This is reflected in a
weak temperature dependence
of the modulation length and wall width, as shown in the top panel of figure \ref{fig:profiles}. At variance with this behavior, for a
large curvature (small $a$ values), the spectrum
of fluctuations is shallow, implying that many modes can be accommodated with a moderate change in energy/temperature. This is again
reflected in the bottom panel
of figure \ref{fig:profiles}, where a sharp change in the magnetization profiles of the bubbles phase is observed as the temperature
is changed.

\begin{figure}[ht!]
\includegraphics[scale=0.34]{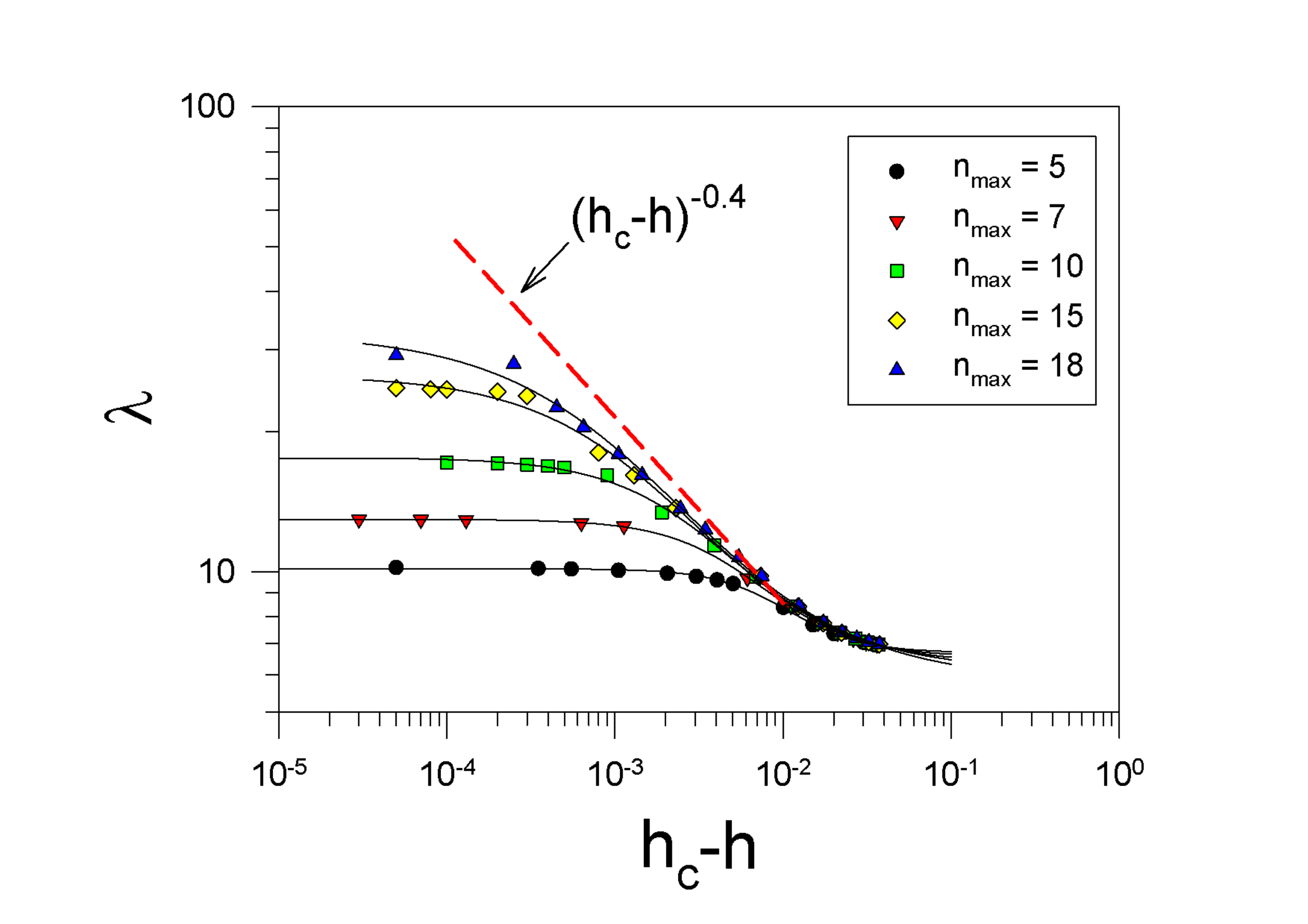}
\caption{ (Color online) Bubbles modulation wavelength $\lambda$ as the transition field $h_c$ is approached from the bubbles phase
at fixed temperature  $T=0.8$ and different values of $n_{max}$,
in the $a=0.2$ model. The continuous lines are fittings using hyperbolic functions in the log-log scale.  }
\label{fig:lambdacrit}
\end{figure}

\begin{figure}[ht!]
\includegraphics[scale=0.33]{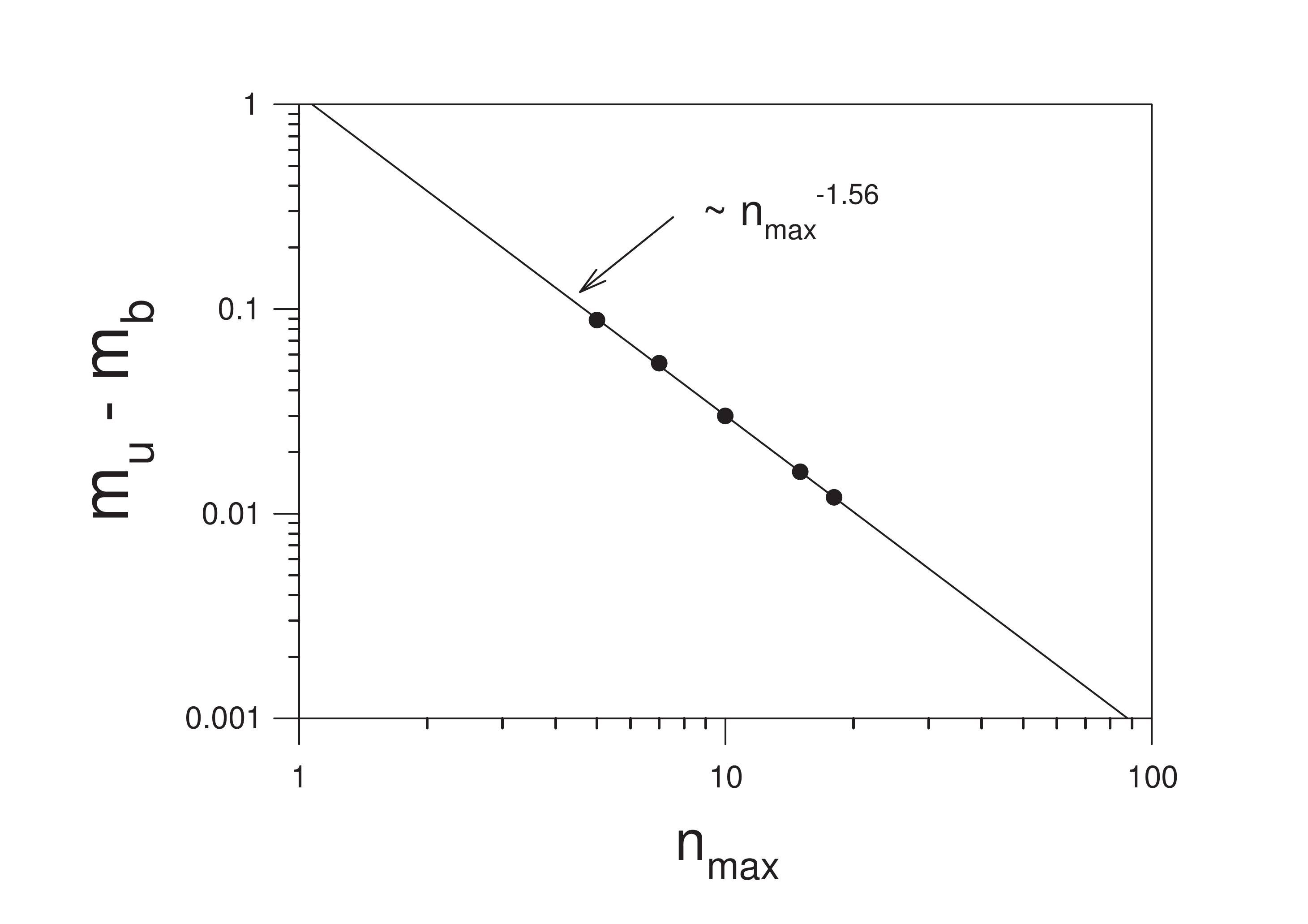}
\caption{(Color online) Magnetization jump at the critical field between the bubbles and uniform phases for $T=0.8$ versus $n_{max}$
in log-log scale, in the $a=0.2$ model. }
\label{fig:deltam}
\end{figure}

Another result of particular interest for the model is the nature of the transition from the bubbles phase to the homogeneous one at
the critical field $h_c$. In figure \ref{fig:lambdacrit} we show the
evolution of the modulation length at fixed temperature $T=0.8$ as the critical field is approached from the bubbles phase for different
values of $n_{max}$, in the $a=0.2$ model. The behavior for $a=4$ and other values of $T$ is qualitatively similar. Previous
works suggested that the critical field lines correspond to first order phase transitions~\cite{CaCaBiSt2011}, with a discontinuous
jump in the modulation
length and magnetization. Nevertheless, as anticipated in an analysis of the stripes solutions for the dipolar frustrated Ising
model~\cite{VeStBi2014}, the transitions
turn out to be continuous in the whole critical line when the limit of large number of modes is reached. Figure \ref{fig:lambdacrit} shows,
for $T=0.8$, that for any finite value
of $n_{max}$ the modulation length $\lambda$ first grows as the field approaches the critical one, but eventually saturates at a finite
value, suggesting a discontinuous
jump at $h_c$. Nevertheless, the saturation value grows itself as $n_{max}$ grows. In order to get an estimation of the asymptotic
behavior, we have fitted the data
(shown in log-log scale) with hyperbolic functions. A scaling analysis of the saturation value of $\lambda$ (not shown) suggests that
it diverges with the power law
$(h_c-h)^{-0.4}$. The dashed line (in red) shows this asymptotic behavior, implying that, in the limit $n_{max}\to \infty$, the
modulation length grows continuously as
$h_c$ is approached and the homogeneous phase emerges as the limit $\lambda \to \infty$. A confirmation of the continuous character of
the phase transition to the
uniform phase comes from an analysis of the size of the jump in the magnetization at the critical field. In figure \ref{fig:deltam} we
show the difference between the magnetizations
of the homogeneous and bubbles ($m_u-m_b$) solutions at $T=0.8$ and $h_c$ versus $n_{max}$. A fit of the five points available in
logarithmic scale shows that the jump goes to zero as $n_{max}\to \infty$
with a power law $n_{max}^{-1.56}$. Preliminary results indicate that the bubbles-paramagnetic transition is continuous in the whole
critical line, determination of critical exponents along the line is left for future work.

\section{Conclusions}
\label{conc}

We introduced a coarse grained model for stripe forming systems (such as ultrathin magnetic films), that generalizes the usual Landau--Ginzburg one. The inclusion of a complete mean field entropic form (instead of its expansion) allowed us to obtain the $(h,T)$ phase
diagram at any temperature, not only close to the critical one. The method is completed by proposing variational modulated solutions,
namely  bubbles and stripes, in the form of appropriately truncated Fourier expansions.

After an appropriated rescaling, the fluctuation spectrum of the model can be characterized by a single parameter, namely, the
inverse curvature at the minimum of the spectrum, $a$. We found that $a$ determines the existence or not of ISB transition.  For
ultrathin magnetic films models, large values of $a$ mean small values of $\delta$, namely, the exchange to dipolar couplings ratio.
We found that for large values of $a$  the phase diagram does not display ISB. This behavior is therefore consistent with previous Monte
Carlo simulation results~\cite{CaCaBiSt2011,DiMu2010} for small values of $\delta$, where no ISB were observed.

For small enough values of $a$ the ISB transition emerges and the phase diagram displays (in the limit $n_{max}\to\infty$) the same
topology observed experimentally in Fe on Cu films\cite{SaLiPo2010,SaRaViPe2010}.   Moreover, when $a$ is small enough, an ISB transition is observed not only between the uniform and the bubbles solution, but also between stripes and bubbles for large enough values of $n_{max}$ (see Fig.\ref{fig:pda02}), an experimentally verified fact\cite{SaLiPo2010}.
Our results  also show the presence of a triple
point between bubbles, stripes and uniform phases for finite values of $n_{max}$. However, such point moves towards lower temperatures as $n_{max}$ increases suggesting that, in the $n_{max}\to\infty$ limit, either it is driven to $T=0$ or it simply disappears. In other words, it would be probably an spurious effect of the finite mode approximation. This last scenario would be consistent with the ground state calculations of Ref.[\onlinecite{SaRaViPe2010}]. On the other hand, our results appear to be consistent   with the existence of a triple point between the three phases at $(T,h)=(T_c,0)$, for any value of $a$, in the sense that,  within our  numerical resolution, both transition lines join at such point and  are independent of $n_{max}$ in its neighborhood.

We observed a clear correlation between the appearance of the ISB transition and the low temperature domain wall behavior at the
modulated phases. In the non reentrant regime, i.e. when $a$ is large, the magnetization profiles in the modulated phases exhibit
extended domain walls whose width (relative to the modulation length) varies very little with the temperature, down to very low values
of $T$. On the contrary, when the ISB transition is present the domain wall width shows a strong variation with temperature, becoming
very sharp as the temperature decreases approaching the ISB. Such phenomenon is completely consistent with the phenomenological scaling
hypothesis stated in Ref.[\onlinecite{PoGoSaBiPeVi2010}], according to which a change in the nature of the domain walls should be enough
to explain the appearance of ISB. The change in magnetization profile is consistent with a shallow spectrum around its minimum
(small values of $a$), since the system can accommodate a large number of modes (necessary condition to develop sharp domain walls)
with a moderate change in energy/temperature. Finally, our results suggest that (at least at the mean field level) the whole transition
line between the modulated and non modulated phases (i.e., bubbles and uniform) is continuous with continuously diverging wave length, consistently
with the results of Ref.[\onlinecite{VeStBi2014}].

\acknowledgments

This work was partially supported by CAPES(Brazil)/SPU(Argentina) through grant PPCP 007/2011, CONICET through grant
PIP 112-201101-00213, SeCyT (Universidad Nacional de C\'ordoba,
Argentina) and CNPq (Brazil).

\subsection[Appendix]{Construction of the bubbles solution}
In this appendix we discuss in some detail how the bubbles solution is constructed and which are its features. In the kind of systems
under study here, the bubble pattern is composed by a triangular regular array of circular bubbles, distributed over a background
of homogeneous magnetization. The bubbles are magnetized contrary to the background in order to minimize the dipolar energy cost
in the free energy functional.

From first principles we know that we can construct our bubble solution $\phi_b({\vec x})$ as a superposition of one dimensional modulations
with the correct set of wave vectors. In this way we can write the solution in the form:
\begin{equation}
    \phi_b({\vec x}) =  \sum_{i=0}^\infty c_{i} \cos (k_{eq}\, {\vec{b}}_i\cdot{\vec x}),
\end{equation}
where the wave vectors $\lbrace\vec{b}_i\rbrace$ are selected as forming a triangular lattice with lattice size equal one. At the
same time, since we have a regular array in which all bubbles are identical, we need to impose some conditions to the Fourier amplitudes
of our solution. This condition implies that all those Fourier amplitudes corresponding to wave vectors  $\vec{b}_i$'s related by symmetry
operations of the triangular lattice are equal.

\begin{figure}[ht!]
\includegraphics[scale=0.33]{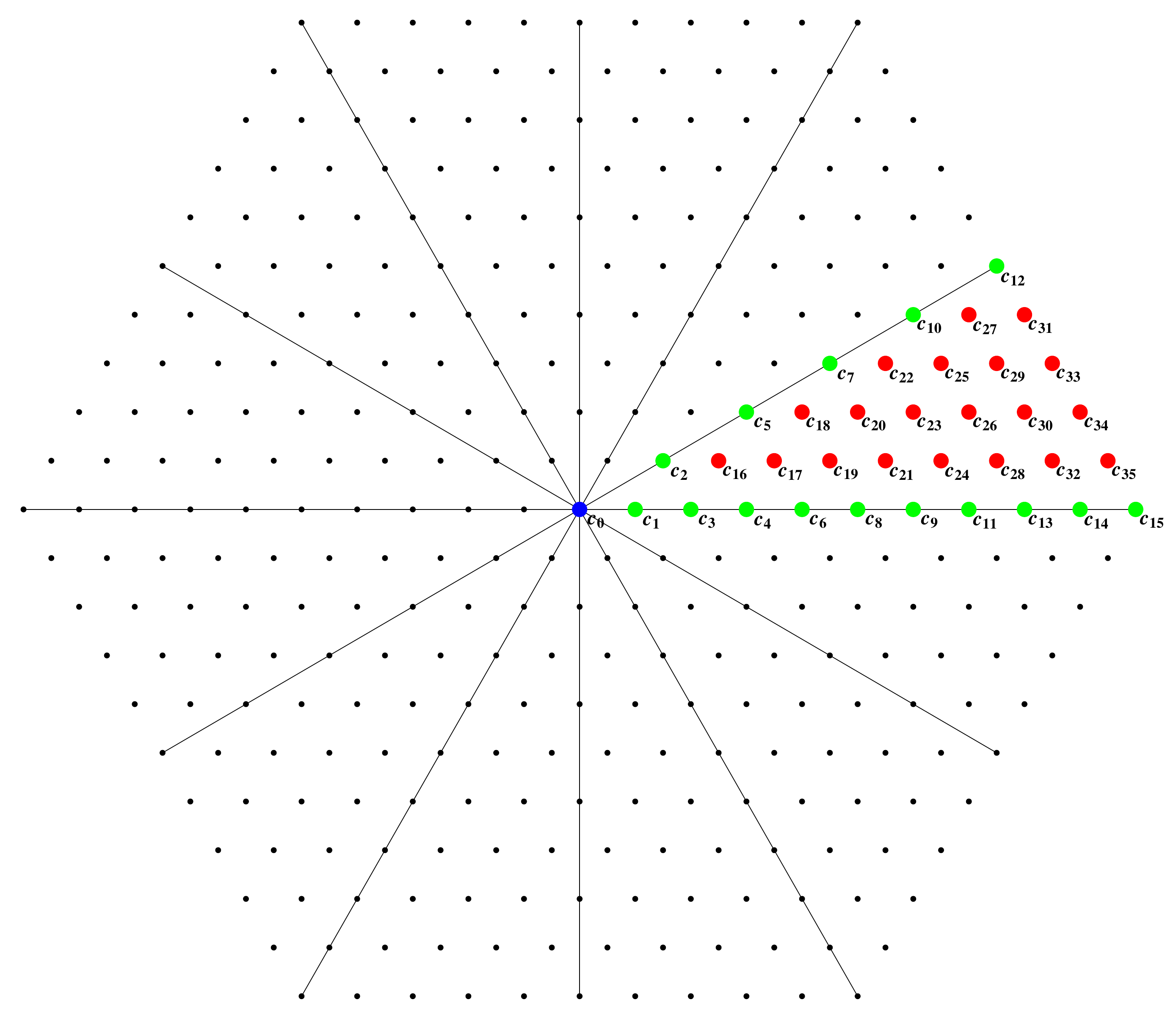}
\caption{(Color online) Lattice of wave vectors for a bubbles pattern considering $n_{\mathrm{max}}=10$. Each dot represents a different
wave vector $\vec{b}_i$ considered to build our solution. }
\label{bbmodes}
\end{figure}

In this way, once we have chosen the maximum number of modes in the principal directions $n_{\mathrm{max}}$, it is automatically defined
how many independent Fourier amplitudes we need to consider. In Figure \ref{bbmodes} we show the case of $n_{\mathrm{max}}=10$. For this
particular choice, of the initial set of $331$ Fourier amplitudes corresponding to the set ${\vec{b}_i}$, after considering the symmetry
arguments we are left with only $35$ independent components, as shown in Fig. \ref{bbmodes} with bigger points.
This reduction in the number of Fourier coefficients greatly simplifies the numerical work. Moreover, as can be observed
from Fig. \ref{bbmodes}, our set of independent Fourier amplitudes can be split in three groups of modes, characterized by different
degeneracies of its components. The first group consists only of the zero mode, which have degeneracy equal to one. The second group
is composed by those modes with angular orientation $\theta=0$ and $\theta=\pi/6$, having a degeneracy of $6$, and the third group
is formed by those vectors with $\theta\in(0,\pi/6)$ which has degeneracy $12$. Taking this fact into account it is possible
to write the original free energy of our solution in terms of the independent Fourier amplitudes.

\bibliographystyle{apsrev4-1}  

%

%

\end{document}